\documentclass[aps,pra,twocolumn,superscriptaddress,showpacs]{revtex4-1}
\def\be{\begin{equation}}
\def\ee{\end{equation}}
\def\bea{\begin{eqnarray}}          
\def\eea{\end{eqnarray}}
\def\bi{\begin{itemize}}
\def\ei{\end{itemize}}

\usepackage{graphicx}

\begin{document}

\title{ 
                           Variational Approach to\\
              Projected Entangled Pair States at Finite Temperature
}

\author{Piotr Czarnik} 
\affiliation{Instytut Fizyki Uniwersytetu Jagiello\'nskiego,
             ul. {\L}ojasiewicza 11, 30-348 Krak\'ow, Poland}

\author{Jacek Dziarmaga} 
\affiliation{Instytut Fizyki Uniwersytetu Jagiello\'nskiego,
             ul. {\L}ojasiewicza 11, 30-348 Krak\'ow, Poland}

\date{ July 8, 2015 }


\begin{abstract}
The projected entangled-pair state (PEPS) ansatz can represent a thermal state in a strongly correlated system. 
We introduce a novel variational algorithm to optimize this tensor network
whose essential ingredient is an auxiliary tree tensor network (TTN).
Since full tensor environment is taken into account, 
with increasing bond dimension the PEPS-TTN ansatz provides the exact Gibbs state. 
Our presentation opens with a 1D version for a matrix product state (MPS-TTN) 
and then generalizes to PEPS-TTN in 2D. 
Benchmark results in the quantum Ising model are presented.
\end{abstract}
\maketitle

\section{ Introduction } 

Since their conception as the density matrix renormalization group (DMRG) \cite{White} --
an algorithm to optimize the matrix product state (MPS) ansatz in 1D \cite{Schollwoeck} -- 
tensor networks proved to be a competitive tool to study strongly correlated quantum 
systems. In the last decade, MPS was generalized to a 2D projected entangled pair state (PEPS) \cite{PEPS} 
and supplemented with an alternative multiscale entanglement renormalization ansatz (MERA) \cite{MERA}. 
These tensor networks avoid the notorious fermionic sign problem \cite{fermions} and PEPS was applied 
to the t-J model of high-$T_c$ superconductivity providing the best results on the market \cite{PEPStJ}. 
The networks -- both MPS \cite{WhiteKagome,CincioVidal} and PEPS \cite{PepsRVB,PepsKagome,PepsJ1J2} -- 
also made some major breakthroughs in search for topological order.
 
Unlike the ground state, thermal states were explored mainly with MPS in 1D \cite{ancillas,WhiteT} 
where they can be prepared by imaginary time evolution. One can follow similar lines in 2D \cite{Czarnik,self} 
-- the PEPS manifold is an efficient representation for Gibbs states \cite{Molnar} -- but accurate evolution is more 
demanding. Alternative direct contractions of the partition function were proposed \cite{ChinaT} but - due to local tensor
update - they are not warranted to become exact with increasing refinement parameter. 
In the following we introduce a variational algorithm to optimize PEPS at finite temperature that both employs 
full tensor update and avoids direct imaginary time evolution.  

\begin{figure}[t]
\includegraphics[width=1.0\columnwidth,clip=true]{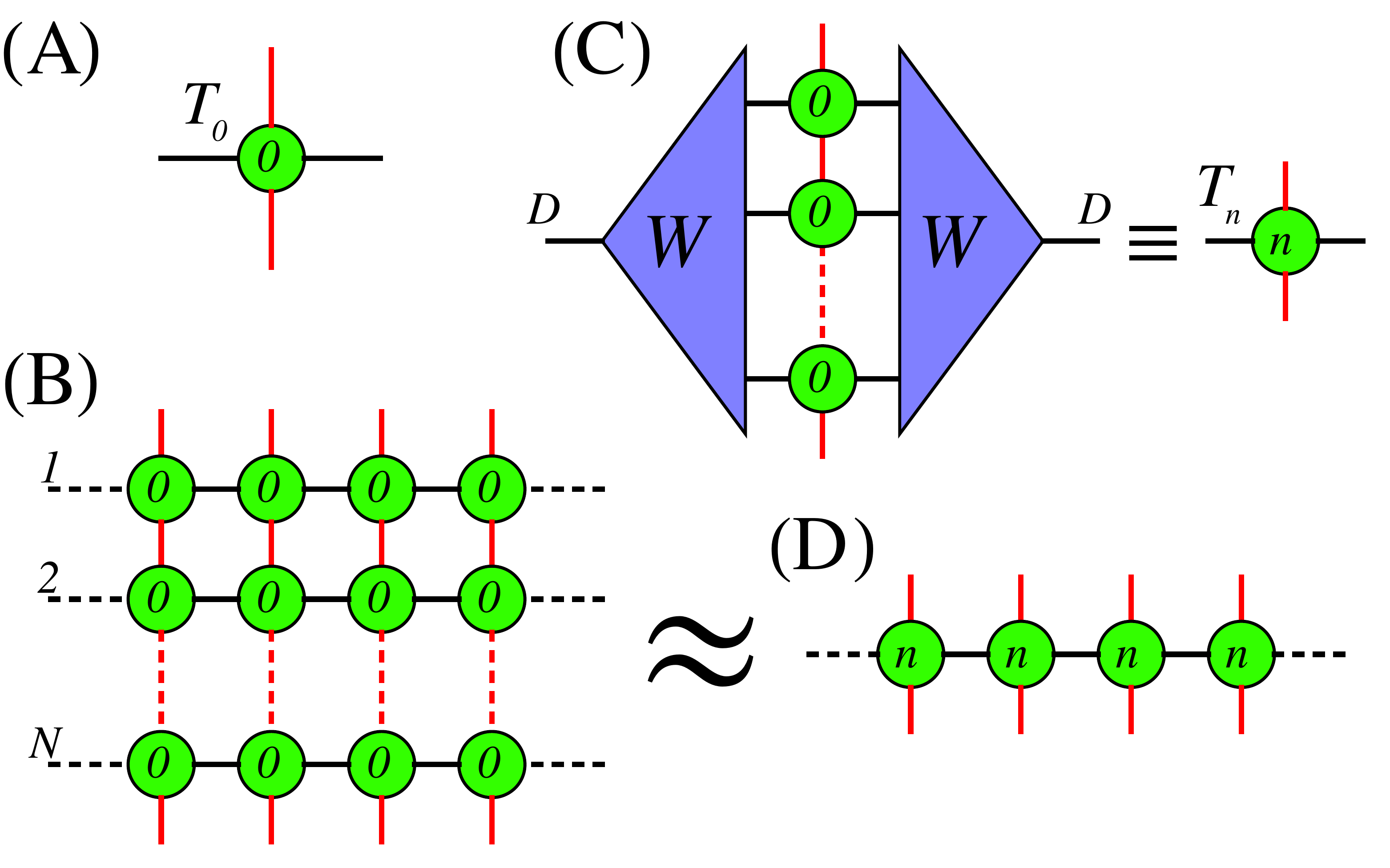}
\caption{ 
In A,
the Trotter tensor $T_0$ in 1D.
This spin operator with (red) spin indices depends on (black) bond indices connecting it with similar operators at the nearest neighbor sites.  
In B,
evolution operator $U(\beta)$ is a product of $N$ time steps represented by horizontal rows of tensors.
Each connecting line is a contracted index.
In C,
at every site the column of elementary $T_0$'s is compressed by an isometry $W$ into one tensor $T_n$ with a bond dimension $D<2^N$.
In D,
evolution operator $U(\beta)$ is approximated by a matrix product operator of the compressed tensors $T_n$.  
}
\label{fig:Trotter1d}
\end{figure}
\section{Purification of thermal states}
\label{sec:purification}
We consider spins on an infinite lattice with a Hamiltonian ${\cal H}$. 
Every spin has states $s=0,...,S-1$ and is accompanied by an ancilla with states $a=0,...,S-1$. 
The enlarged ``spin+ancilla'' space is spanned by states $\prod_m |s_m,a_m\rangle$
where $m$ numbers lattice sites. 
The Gibbs state of spins at an inverse temperature $\beta$ is obtained from its purification
$|\psi(\beta)\rangle$ in the enlarged space, 
\be 
\rho(\beta) ~\propto~
{\rm Tr}_{\rm ancillas}|\psi(\beta)\rangle\langle\psi(\beta)|.
\label{rhobeta}
\ee
At $\beta=0$ we choose a product over lattice sites,
\be 
|\psi(0)\rangle = \prod_m ~\sum_{s=0}^{S-1} |s_m,s_m\rangle ,
\label{psi0}
\ee
to initialize the imaginary time evolution,
\be 
|\psi(\beta)\rangle~=~
e^{-\frac12\beta{\cal H}}~|\psi(0)\rangle~\equiv~
U(\beta)~|\psi(0)\rangle.
\label{psibeta}
\ee
Here the Hermitian $U(\beta)=e^{-\frac12\beta{\cal H}}$ acts in the Hilbert space of spins. With the initial state (\ref{psi0}) the trace in Eq. (\ref{rhobeta}) yields
\be 
\rho(\beta) ~\propto~ U(\beta) ~ U(\beta).
\label{UU} 
\ee
In the following the operator $U(\beta)$ will be represented by a projected entangled-pair operator (PEPO). Thanks to the simple Eq. (\ref{psi0})
equation \ref{psibeta}) translates between the PEPO and a PEPS for the purification $|\psi(\beta)\rangle$ in a trivial way: to obtain the PEPO it 
is enough to rebrand the PEPS's ancilla indices as spin indices.

\section{Quantum Ising model}
\label{sec:ising}
We proceed with the quantum Ising model:
\be 
{\cal H} ~=~- 
\sum_{\langle m,m'\rangle}Z_mZ_{m'}
-h \sum_m X_m.
\label{calH}
\ee
Here $Z,X$ are Pauli matrices and $h$ is a transverse field.
In 1D, the model has a quantum critical point at $h=1$ that becomes a crossover at finite temperature.
On a 2D square lattice,
there is a ferromagnetic phase for small $h$ and large $\beta$. 
At zero temperature the quantum critical point is $h_0=3.044$, see Ref. \cite{hc},
and at $h=0$ the Onsager's critical point is $\beta_0=-\ln(\sqrt{2}-1)/2=0.441$. 
Lattice symmetries are not broken in any phase.

\begin{figure}[t]
\includegraphics[width=1.0\columnwidth,clip=true]{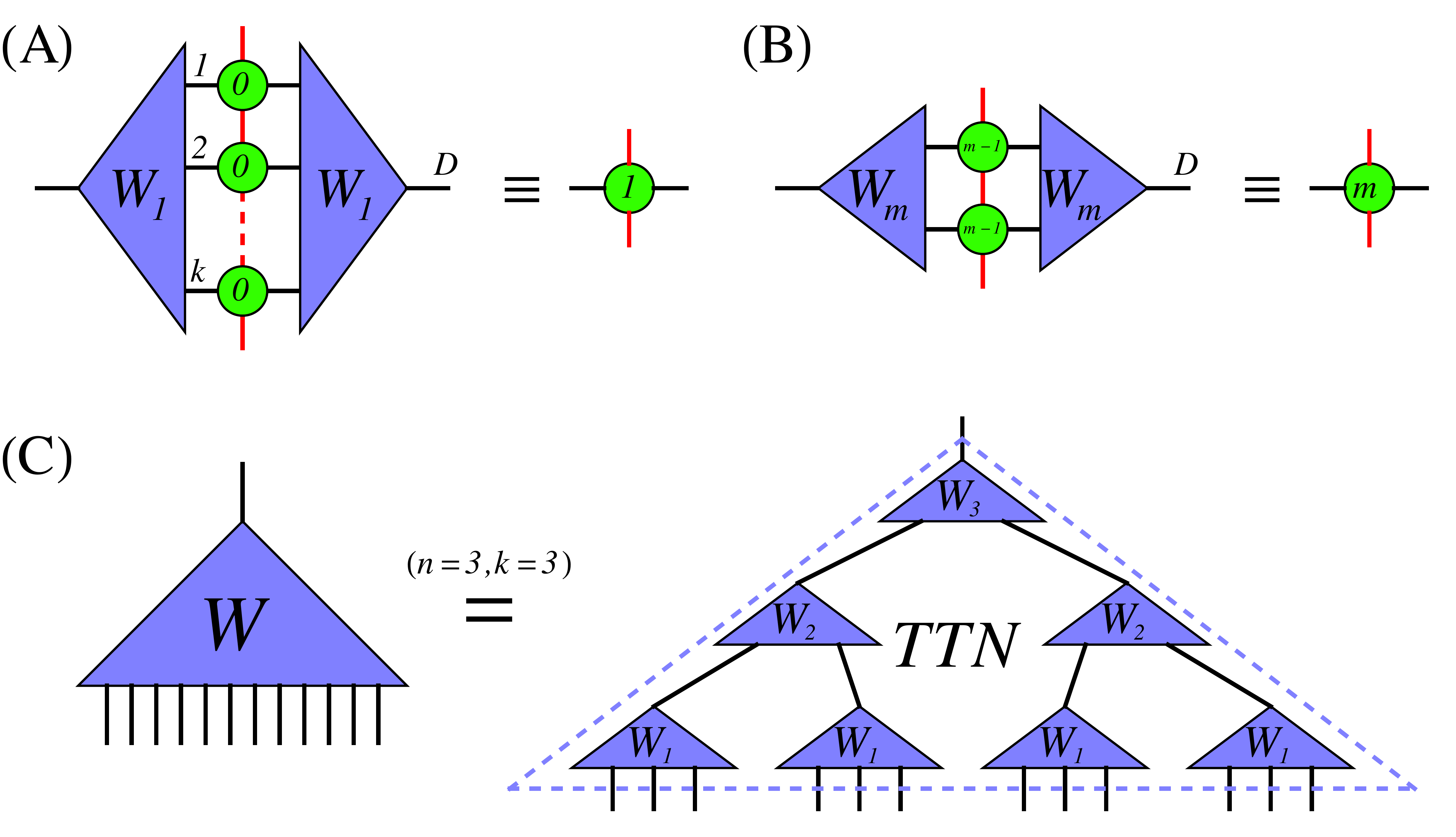}
\caption{ 
The huge compression in Fig. \ref{fig:Trotter1d}C is split into $n$ steps.
In A,
as a first step $k$ elementary Trotter tensors $T_0$ are compressed by isometry $W_1$ into tensor $T_1$ with a bond dimension $D\leq2^k$.
In B, 
the first step is followed by iterative compressions $T_{m-1}\to T_m$ preserving the bond dimension $D$.
In C,
the final $T_n$ is the same as if the huge isometry $W$ were a tree tensor network (TTN) with $n$ layers of isometries: $W_1,...,W_n$.
Here we show an example with $k=3$ and $n=3$.
}
\label{fig:TTN1D}
\end{figure}
\begin{figure}[t]
\includegraphics[width=1.0\columnwidth,clip=true]{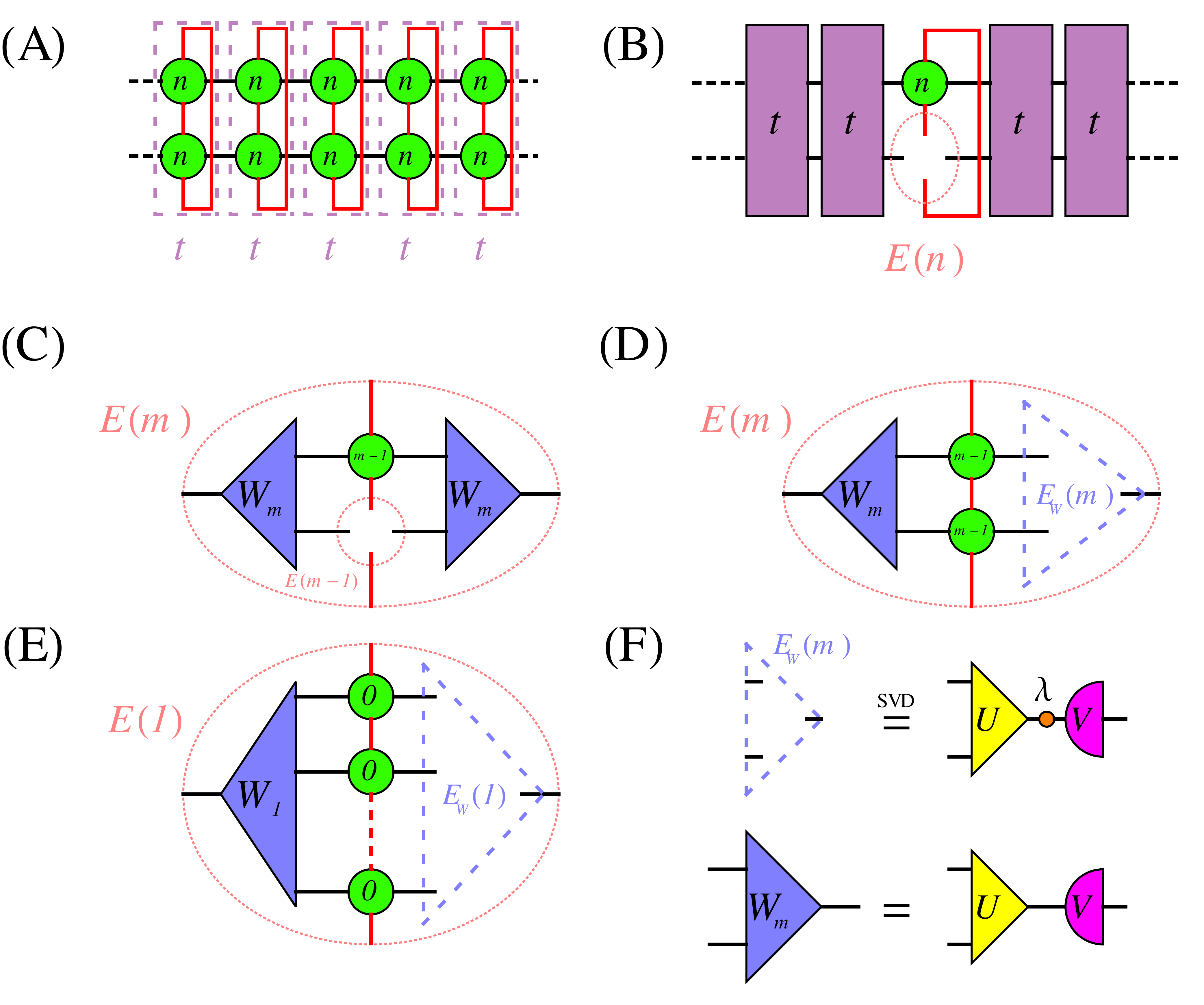}
\caption{ 
In A,
the partition function $Z=\langle\psi(\beta)|\psi(\beta)\rangle$ is a product of transfer matrices $t$.
In B,
the same as in A but with one of the tensors $T_n$ removed. 
This diagram is the tensor environment $E(n)$ for $T_n$.
In C,
a lower-level environment $E(m-1)$ for $T_{m-1}$ is obtained from $E(m)$. 
In D,
environment $E_W(m)$ for isometry $W_m$ is obtained from $E(m)$.
In E,
environment $E_W(1)$ is obtained from $E(1)$.
In F,
isometric environment is subject to singular value decomposition, $E_W(m)=U\lambda V^\dag$,
with $D$ non-zero singular values $\lambda$, a $D^2\times D$ isometry $U$, and a $D\times D$ unitary $V$.
The isometry is updated as $W_m=UV^\dag$ to maximize the figure of merit $Z={\rm Tr~} E_W(m) W_m^\dag$. 
}
\label{fig:Env1D}
\end{figure}

\section{Suzuki-Trotter decomposition}
\label{sec:ST} 
We define gates
\be 
U_{ZZ}(\beta) \equiv \prod_{\langle m,m'\rangle}e^{\frac{\beta}{2}Z_mZ_{m'}},~
U_X(\beta)    \equiv \prod_m e^{\frac{\beta}{2}hX_m}.
\ee
In the second-order Suzuki-Trotter decomposition a small time step is approximated by their product
\be
U(d\beta)\approx U_X(d\beta/2)U_{ZZ}(d\beta)U_X(d\beta/2).
\label{U}
\ee
In order to rearrange $U(\beta)$ as a tensor network,
at each bond we make a singular value decomposition
\be 
e^{\frac{d\beta}{2}Z_mZ_{m'}}=
\sum_{b=0,1}
z_{m,b}~
z_{m',b}.
\ee
Here $b$ is a bond index and 
$z_{m,b}\equiv\sqrt{\Lambda_b}(Z_m)^b$ with the singular values 
$\Lambda_0=\cosh\frac{d\beta}{2}$ and $\Lambda_1=\sinh\frac{d\beta}{2}$. 
Now we can write 
\bea
U(d\beta)=
\sum_{\{b\}}
\prod_m
\left(
e^{\frac{d\beta}{4}hX_m}
\prod_{m'}
z_{m,b_{\langle m,m'\rangle}}~
e^{\frac{d\beta}{4}hX_m}
\right).
\eea
Here $\{b\}$ is a set of all bond indices $b_{\langle m,m'\rangle}$.
The brackets enclose an elementary Trotter tensor $T_0$ at site $m$.
It is a spin operator depending on bond indices connecting its site with its nearest neighbors. 
Its 1D version is shown in Fig. \ref{fig:Trotter1d}A. 

The evolution is a product of $N$ small time steps,
\be 
U(\beta)=\left[U(d\beta)\right]^N.
\ee
For pedagogical reasons, we begin with a 1D version of this product in Fig. \ref{fig:Trotter1d}B, 
where each row $1...N$ is the elementary time step $U(d\beta)$ and 
each column is a site in the 1D chain. 
We need an efficient algorithm to make this huge sum.

\section{MPS-TTN algorithm in 1D}
\label{sec:MPS}
The general idea is to compress first the $N$ elementary tensors $T_0$ at every site into a single tensor $T_n$ as in Fig. \ref{fig:Trotter1d}C 
and then to contract the compressed $T_n$'s horizontally as in Fig. \ref{fig:Trotter1d}D. 
Here $W$ is an isometry from the auxiliary Hilbert space spanned by $2^N$ possible values of $N$ bond indices to its $D$-dimensional subspace. 
The bond dimension $D$ is a refinement parameter: with increasing $D$ results should become numerically exact. 

Furthermore, 
instead of making the huge compression $W$ all at once -- that is all but possible -- we split it into $n$ steps as in Figs. \ref{fig:TTN1D}A, \ref{fig:TTN1D}B.
The final $T_n$ is the same as if $W$ were a tree tensor network (TTN) \cite{TTN} with $n$ layers of isometries $W_1,...,W_n$, see the example in Fig. \ref{fig:TTN1D}C. 
Notice that the number of different isometries is only logarithmic in the number $N=2^{n-1}k$ of time steps $d\beta$:
\be 
n ~=~ 1+\log_2\frac{N}{k} ~\sim~ \log N ~=~ \log\frac{\beta}{d\beta}.
\ee
Low temperatures or near-infinitesimal $d\beta$ required in precise applications can be achieved with a marginal logarithmic cost.  
In the following we describe how this variational TTN ansatz can be optimized in an efficient way.

The optimization aims to maximize the partition function
\be 
Z\left[W\right]=e^{-\beta F}={\rm Tr}~e^{-\beta{\cal H}}={\rm Tr~} U(\beta) U(\beta)
\ee
with respect to the isometries $W_1,...,W_n$.
This figure of merit has to be maximized in order to minimize the Gibbs free energy $F$.
With each $U(\beta)$ in Eq. (\ref{UU}) represented by the diagram in Fig. \ref{fig:Trotter1d}D,
$Z$ becomes the tensor network in Fig. \ref{fig:Env1D}A. 

In order to maximize $Z$ with respect to $W_m$ we need a gradient $\partial Z/\partial W_m$.
In principle, 
the gradient is an infinite sum of derivatives with respect to every $W_m$ in $Z$ but,
thanks to the symmetries,
all these derivatives are the same and equal to a tensor environment $E_W(m)$ of $W_m$.
The environment is the partition function in Fig. \ref{fig:Env1D}A, 
but with one tensor $W_m$ removed.

The environment can be computed efficiently by an algorithm made of the procedures depicted in Figs.\ref{fig:Env1D}B-E.
In Fig. \ref{fig:Env1D}B we show an environment $E(n)$ for the compressed tensor $T_n$.
This environment is the partition function in Fig.\ref{fig:Env1D}A, but with one tensor $T_n$ removed.
It can be calculated with the standard transfer-matrix techniques applied to the transfer matrix $t$.
Figure \ref{fig:Env1D}C shows how to compute an environment $E(m-1)$ for tensor $T_{m-1}$ from a higher-level environment $E(m)$ for $T_m$.
Repeating this procedure for $m=n,...,2$ 
we could obtain all Trotter tensors' environments from $E(n-1)$ down to $E(1)$ in one go.
However, after each $E(m)$ is obtained we pause to calculate an environment $E_W(m)$ for the isometry $W_m$ by contracting 
the diagram in Fig. \ref{fig:Env1D}D or its variant in Fig. \ref{fig:Env1D}E for $m=1$.
This environment is used immediately to optimize $W_m$ by the SVD technique depicted 
in Fig. \ref{fig:Env1D}F. With the updated $W_m$ the algorithm proceeds to calculation of $E(m-1)$ and so forth all the way 
down to $E(1)$ and $W_1$ when the down optimization sweep is finally completed. The end of the down sweep is the beginning 
of an up sweep whose first step is calculation of $T_1$ as in Fig. \ref{fig:TTN1D}A. With $T_1$ we calculate $E_W(2)$ 
contracting the diagram in Figure \ref{fig:Env1D}D and then immediately update $W_2$ before proceeding to calculation of 
$T_2$. This procedures are repeated all the way up to $W_n$ and $T_n$ when the up sweep is completed. Before the next 
down sweep the environment $E(n)$ in Fig. \ref{fig:Env1D}B is updated. The whole optimization procedure is 
repeated until convergence.

In summary, the variational TTN is optimized by repeated up- and down-sweeps. 
The up-sweep is a sequence
\bea
T_0 \to E_W(1) \to T_1 \to .... \to E_W(n) \to T_n \to E(n) \nonumber
\eea
followed by the down sweep
\bea 
E(n) \to E_W(n) \to E(n-1) \to .... \to E(1) \to E_W(1). \nonumber
\eea
Each environment $E_W(m)$ is used immediately to update $W_m$. 
Progress of the optimization is monitored with a set of figures of merit $Z_m={\rm Tr~}E_W(m)W_m^\dag$. 
When all $Z_m$ become the same within presumed numerical accuracy, 
then the isometries are accepted as converged. 
The numerical cost of the 1D algorithm scales like $(D^2)^3$, 
where $D^2$ is the bond dimension of the transfer matrix $t$. 

\begin{figure}[t]
\includegraphics[width=1.0\columnwidth,clip=true]{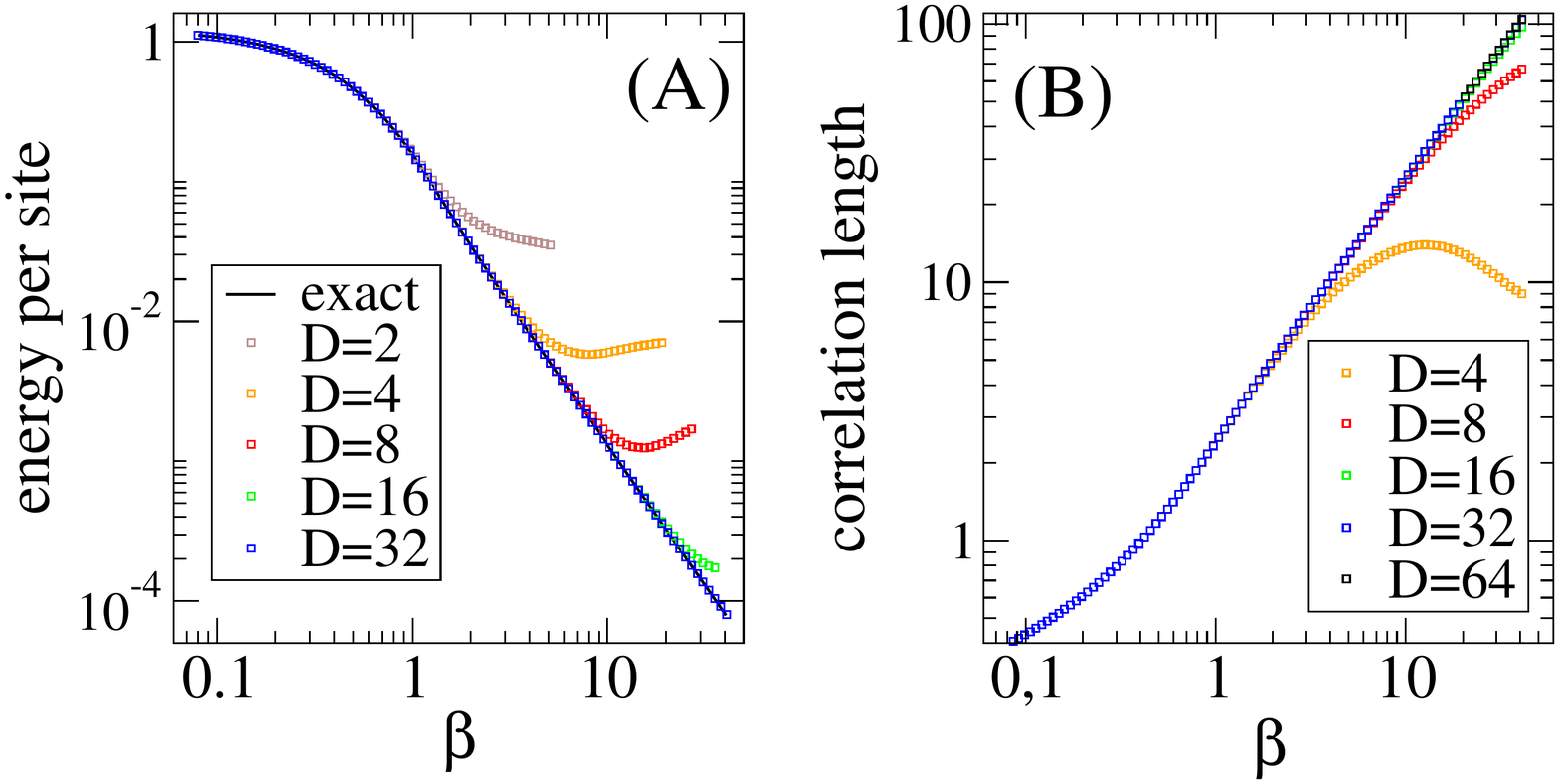}
\vspace{-2.3cm}
\caption{ 
In A,
energy per site of the quantum Ising chain at the critical field $h=1$
in function of $\beta$ for different bond dimensions $D$.
For larger $D$ the results remain accurate down to lower temperatures
that are closer to the quantum critical point.
In B,
the correlation length in the tail of the ferromagnetic correlator (\ref{CR1D}). 
In both A and B, 
the number of isometry layers $n=11$, 
the bond dimension $D=2^k=2,...,64$, 
and the elementary time step in the Suzuki-Trotter decomposition $d\beta=\beta/(2^{n-1}k)$.     
}
\label{fig:U1d}
\end{figure}

\begin{figure}[t]
\vspace{-1.1cm}
\includegraphics[width=1.15\columnwidth,clip=true]{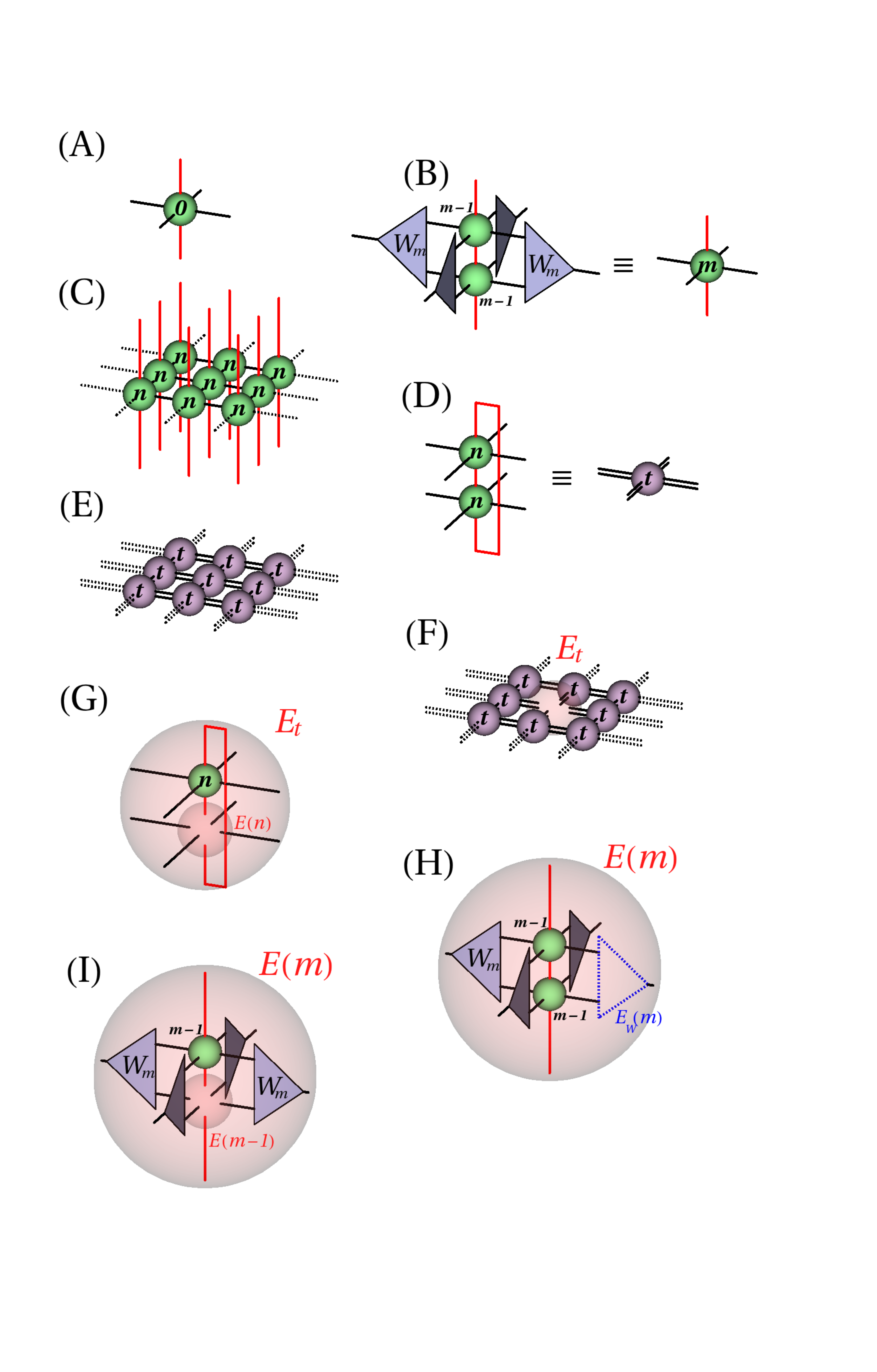}
\vspace{-2.2cm}
\caption{ 
In A,
the elementary Trotter tensor $T_0$ on a square lattice.
In B,
compression of two tensors $T_{m-1}$ into one $T_m$ with four isometries $W_m$.
In C,
tensors $T_n$ are contracted into a PEPS network for the evolution $U(\beta)$.
In D,
a contraction of two tensors $T_n$ makes a transfer tensor $t$.
In E,
contraction of transfer tensors is the partition function $Z$.
In F,
the partition function with one of the transfer tensors removed is a tensor environment $E_t$ for the removed tensor $t$.
We enclose the ends of the free bonds by a transparent reddish sphere. Each free bond has dimension $D^2$.
In G,
when the environment/sphere $E_t$ is filled and contracted with one Trotter tensor $T_n$ we obtain tensor environment $E(n)$ for $T_n$. 
This environment is the partition function with one $T_n$ removed.
In H,
environment $E(m)$ filled and contracted with two $T_{m-1}$'s and three $W_m$'s makes environment $E_W(m)$ for isometry $W_m$. 
In I,
$E(m-1)$ is obtained from $E(m)$.
}
\label{fig:peps}
\end{figure}

\section{Benchmark results in 1D}
\label{sec:MPSbench}

We applied the algorithm to scan thermal states at the critical field $h=1$ up to $\beta\approx40$,
see Fig. \ref{fig:U1d} where we show the energy per site and the correlation length $\xi$ in
the ferromagnetic correlator
\be 
C_R=\langle Z_{x} Z_{x+R} \rangle-\langle Z_{x}\rangle\langle Z_{x+R}\rangle
\label{CR1D}
\ee
with an exponential tail
$
C_R \propto e^{-R/\xi}
$
for large $R$. We find that for increasing bond dimension $D=2^k$ results remain accurate
down to lower temperatures, reaching closer to the quantum critical point.

The number of TTN layers was fixed at $n=11$ to achieve small enough $d\beta$ for the results
to be converged in $d\beta$ all the way down to $\beta\approx40$. Increasing $n$ up to $20$
did not affect stability of the algorithm. The plots in Fig. \ref{fig:U1d} are scans in 
increasing $\beta$ in the sense that isometries converged for one $\beta$ were used as 
the initial isometries for the next $\beta+\delta\beta$. This recycling reduced the number of
optimization sweeps necessary for each $\beta$ to $10...100$. This number increased with
$\beta$ as the quantum critical point was approached.

\section{PEPS-TTN algorithm in 2D}
\label{sec:PEPS}
The algorithm can be generalized to a 2D square lattice as summarized in Figure \ref{fig:peps}.
The elementary Trotter tensor $T_0$ now has four bond indices to be contracted with its four nearest neighbors,
see Fig. \ref{fig:peps}A. Accordingly, each compression of two Trotter tensors $T_{m-1}$ into 
a higher level tensor $T_m$ is done with four isometries $W_m$, see Fig. \ref{fig:peps}B.
After all the isometric compressions are completed, the operator $U(\beta)$ can be represented by 
the PEPO in Fig. \ref{fig:peps}C with tensors $T_n$. A contraction of two $T_n$'s makes a transfer tensor 
$t$, see Fig. \ref{fig:peps}D. The infinite network of transfer tensors in Fig. \ref{fig:peps}E represents 
the partition function $Z$. This 2D network replaces the 1D chain of transfer matrices in Fig. \ref{fig:Env1D}A. 
When one of the transfer tensors in Fig. \ref{fig:peps}E is removed from the partition function, 
then we obtain its tensor environment $E_t$ represented by the network in Fig. \ref{fig:peps}F. 

Unlike its analogue in 1D, where the exact transfer-matrix techniques can be used, this environment 
requires more sophisticated technology. In this paper we use the symmetric version of the corner matrix 
renormalization (CMR) \cite{CMR} whose description and discussion is delegated to appendix \ref{CMR}. 
Numerically it is the most expensive part of the algorithm with an additional refinement parameter of 
its own: the environmental bond dimension $M$. 

Once $E_t$ is converged, we can continue with the main loop of the algorithm. In Fig. \ref{fig:peps}G 
$E_t$ is contracted with $T_n$ to yield an environment for $T_n$ that we call briefly $E(n)$. 
With $E(n)$ the down optimization sweep begins that proceeds as follows. From $E(n)$ we obtain an environment 
$E_W(n)$ for the isometry $W_n$ as in Fig. \ref{fig:peps}H. Once calculated, this environment is used immediately 
to update $W_n$ as in Fig. \ref{fig:Env1D}F. With the updated $W_n$ we proceed to calculate the environment 
$E(n-1)$ for $T_{n-1}$ as in Fig. \ref{fig:peps}I and then $E_W(n-1)$ to update $W_{n-1}$. The same procedure
repeats all the way down to $W_1$. 

After all the isometries were optimized down to $W_1$, the upwards optimization sweep begins. It has $n$ steps. 
In the $m$-th step two tensors $T_{m-1}$ and the environment $E(m)$ -- that was calculated before during the 
down sweep -- are contracted to obtain $E_W(m)$, see Fig. \ref{fig:peps}H. With this environment $W_m$ is updated 
immediately and then used to compress two $T_{m-1}$ into one $T_m$ as in Fig. \ref{fig:peps}B. This basic step
is repeated all the way up to $T_n$.  

Just as in 1D, the above description can be briefly summarized as follows. The up-sweep is a sequence
$$
T_0 \to E_W(1) \to T_1 \to .... \to T_{n-1} \to E_W(n) \to T_n~
$$
completed with the CMR procedure
$$ 
T_n\to E_t\to E(n)~.
$$
that is also the starting point for the down-sweep,
$$
E(n) \to E_W(n) \to E(n-1) \to .... \to E(1) \to E_W(1) ~.
$$
Here, both in the up- and down-sweeps, each $E_W(m)$ is used immediately to update $W_m$. The whole up-down cycle 
is repeated until convergence. 

The numerical cost of the procedures in panels B,D,G,H,I of Fig. \ref{fig:peps} scales formally like $(D^2)^4$, 
where $D^2$ is the bond dimension of the transfer tensor. The exponent is steeper than in 1D, but a much smaller 
$D$ is typically needed in 2D. For instance, unlike in 1D, $D$ can be finite at a critical point. However, the actual 
bottleneck is the CMR in appendix \ref{CMR}, whose cost scales like $(D^2)^3M^3$. In non-symmetric versions of 
CMR \cite{CMR} this formal cost can be cut down to $(D^2)^2M^3$. The parameter $M$ is 
expected to diverge at a critical point together with a diverging correlation length but, even at criticality, 
local observables and correlations at increasingly long distance can be converged with increasing $M$. 

\begin{figure}[t]
\includegraphics[width=1.05\columnwidth,clip=true]{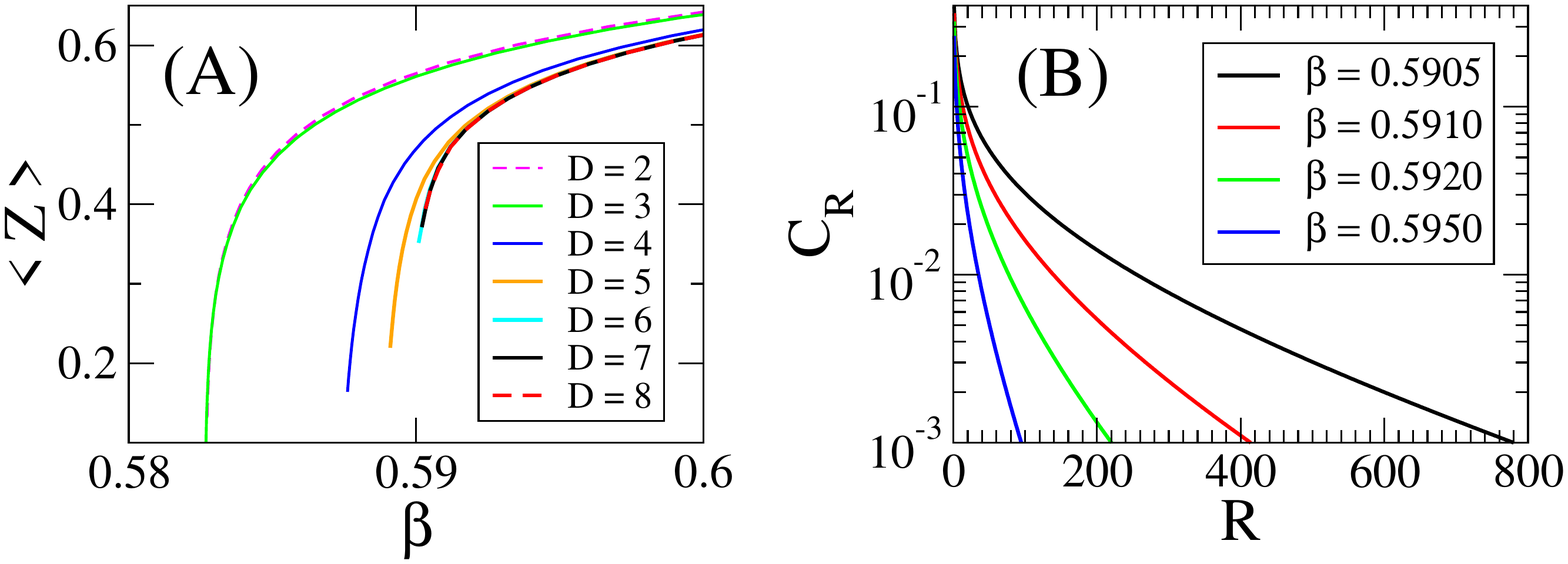}
\vspace{-3.3cm}
\caption{ 
In A,
spontaneous magnetization $\langle Z \rangle$ in the ferromagnetic phase
at the transverse field $h=\frac23 h_0$. Different colors correspond to 
bond dimensions $D=2...8$. Plots for $D=6,7,8$ collapse demonstrating convergence 
for $D\geq6$.
In B, 
the ferromagnetic correlator $C_R$ in Eq. (\ref{CR}) for $D=6$ at $h=\frac23 h_0$ 
and different $\beta$ in the ferromagnetic phase. Its correlation length diverges 
closer to criticality requiring a diverging $M$. For $M=68$ the longest length 
achieved is $\xi=290$ lattice sites.
}
\label{fig:U2d}
\end{figure}

\begin{figure}[t]
\includegraphics[width=1.05\columnwidth,clip=true]{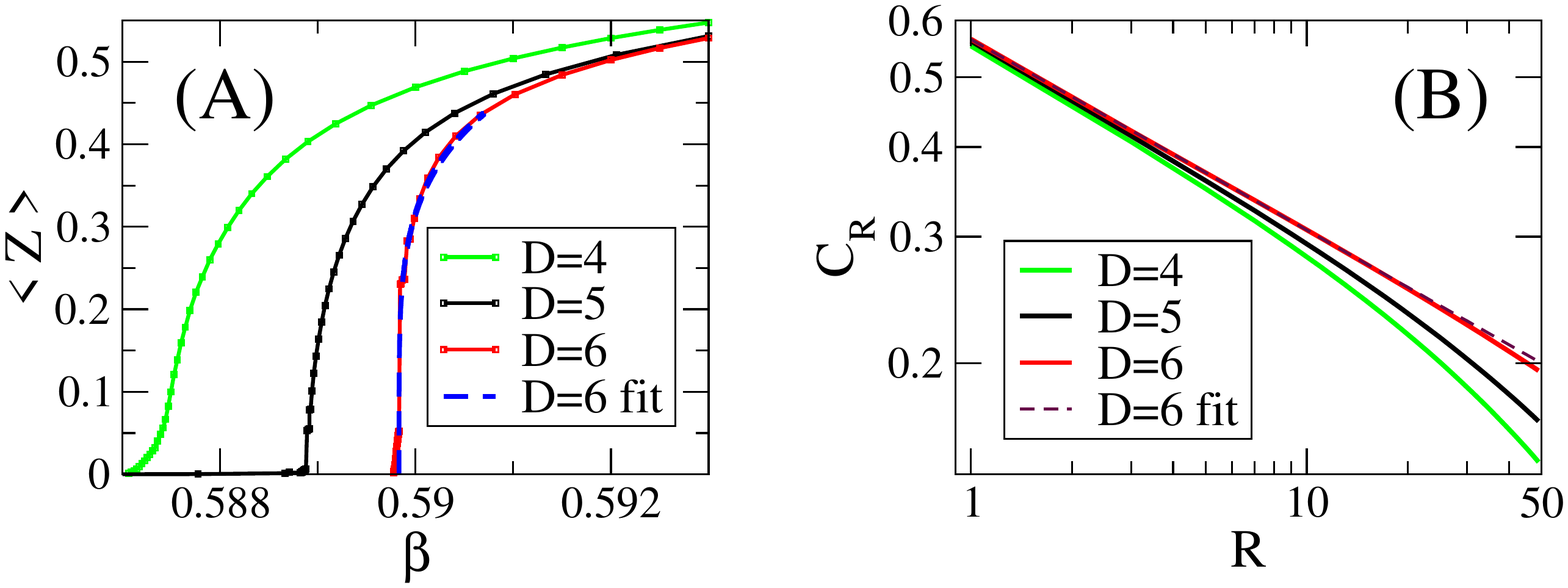}
\vspace{-3.3cm}
\caption{ 
In A,
spontaneous magnetization $\langle Z\rangle$ near the critical point.
With increasing bond dimension $D$ the plots become ``critical'' enough to make
a fit $\langle Z\rangle\propto \left(\beta-\beta_c\right)^{\beta'}$ 
with a critical-point $\beta_c=0.5898$ and a critical exponent $\beta'=0.198$.
In B,
the ferromagnetic correlation function (\ref{CR}) for $D=6$ at $\beta_c$.
In the intermediate range $1<R<30$ the correlator is fitted well by 
the power law $C_R\propto R^{-\eta}$ with the critical exponent $\eta=0.265$. 
}
\label{fig:U2dcrit}
\end{figure}

\section{Benchmark results in 2D}
\label{sec:PEPSbench}

We applied the PEPS-TTN algorithm to scan the ferromagnetic phase at the transverse field 
\be 
h=\frac23h_0.
\ee 
Results in Fig. \ref{fig:U2d} show convergence in the bond dimension (for $D\geq6$). 
For each $D$ they are converged in the environmental bond dimension $M$. 
The plots do not extend all the way down to the critical point, 
because the diverging correlation length would require a diverging environmental bond dimension $M$ 
to obtain a Gibbs state fully converged in $M$.
Nevertheless,
even with a limited $M\leq68$ we could get close enough to the critical point to obtain
a converged ferromagnetic correlation function
\be 
C_R=\langle Z_{x,y} Z_{x+R,y} \rangle-\langle Z_{x,y}\rangle\langle Z_{x+R,y}\rangle
\label{CR}
\ee
with a long correlation length up to $\xi=290$ in its exponential tail
$
C_R \propto e^{-R/\xi}
$
for large $R$. 

The plots in Fig. \ref{fig:U2d} are fully converged in $M$, 
but they terminate before the critical point provoking a natural question 
what, if anything, can be achieved closer to criticality.
Hence we pushed our computations closer, 
accepting the fact that the ansatz cannot be fully converged there. 
Results are shown in Fig. \ref{fig:U2dcrit}. 
The spontaneous magnetization for $D=6$ and $M=35$ allows a power law fit that gives the order parameter exponent 
$\beta'=0.198$ (the textbook exponent $\beta$ distinguished here by a prime from the inverse temperature) and the 
critical point $\beta_c=0.5898$. 
The correlation function at $\beta_c$ has an exponential tail that is not converged in $M$ -- 
its range increases with $M$ --
but at an intermediate range $1<R<30$ the correlator has the correct form $C_R\propto R^{-\eta}$ with the critical 
exponent $\eta=0.265$. 

To give an idea about practical effectiveness of the algorithm, 
it would take 1 day on a 4-core laptop computer to reproduce Fig. \ref{fig:U2d} and,
since the convergence is slower near criticality,
2-3 more days for Fig. \ref{fig:U2dcrit}. 
Our TTN had $n=6$ layers of isometries with $k=5$ in the bottom layer.
The computation time was checked to be practically independent of $n$, 
as it is CMR that is the actual bottle-neck and not the isometry optimizations. 
For random initial conditions, 
the number of up- and down- optimization sweeps necessary to reach convergence was $\simeq100$. 
For a scan in $\beta$ -- when tensors converged for one $\beta$ were
recycled as initial tensors at a near $\beta-\delta\beta$ -- 
the number of sweeps was typically $\simeq10$. Both numbers increased towards the critical point.

\section{Comparison with direct imaginary time evolution}
\label{sec:comparison} 

\begin{figure}[t]
\includegraphics[width=1.0\columnwidth,clip=true]{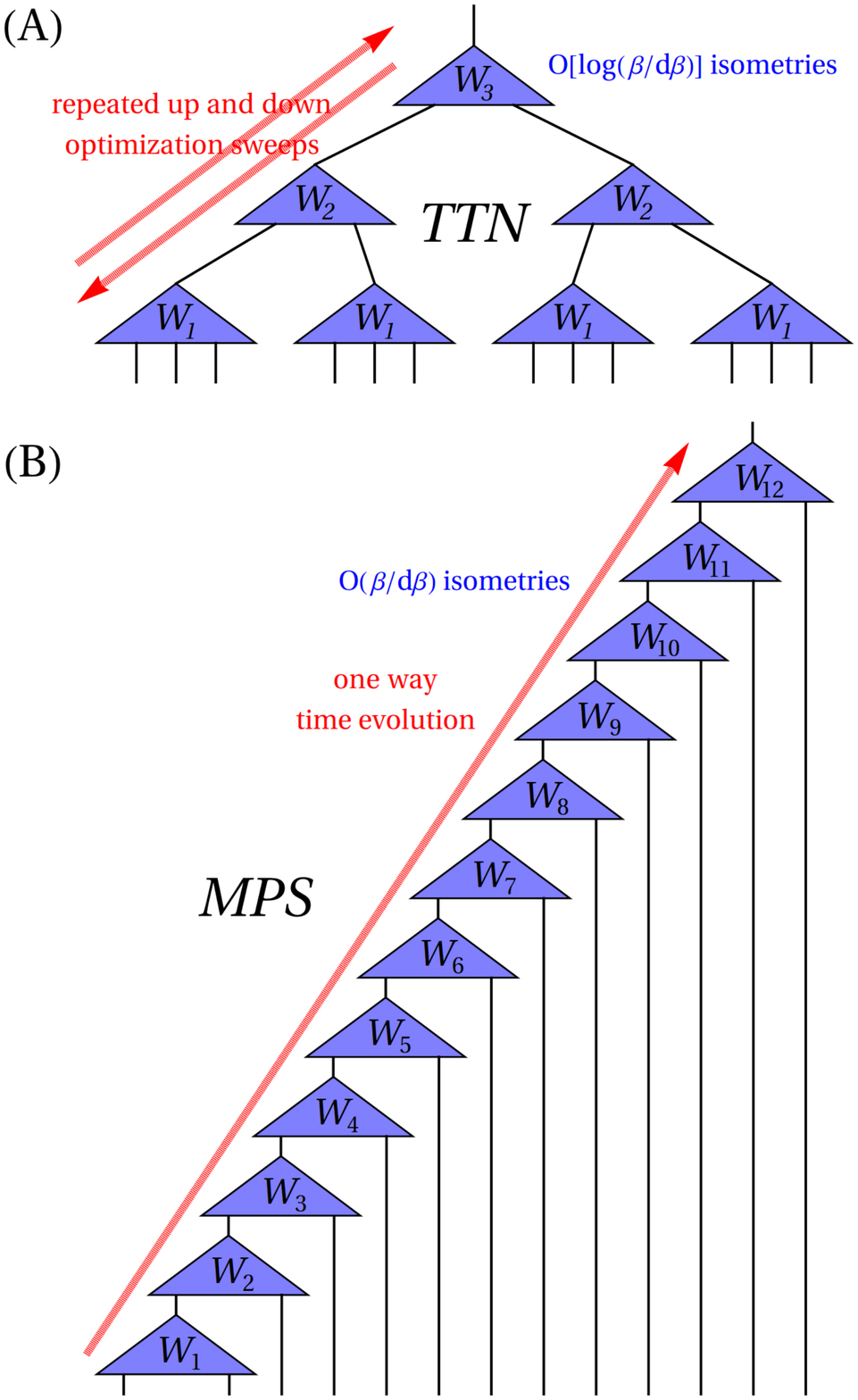}
\caption{ 
Comparison between the present variational method in A and the direct imaginary time evolution in B. 
The TTN ansatz is exponentially more compact than MPS since the number of different isometries is only logarithmic 
in the number of time steps. On top of this, in the present variational method the isometries in TTN
are optimized by repeated up- and down-sweeps to provide only the most accurate state at the final $\beta$,
while in the time evolution all intermediate states between $0$ and the final $\beta$ need to be accurate
compromising the accuracy of the final one. 
}
\label{fig:ImTime}
\end{figure}
\begin{figure}[t]
\includegraphics[width=0.8\columnwidth,clip=true]{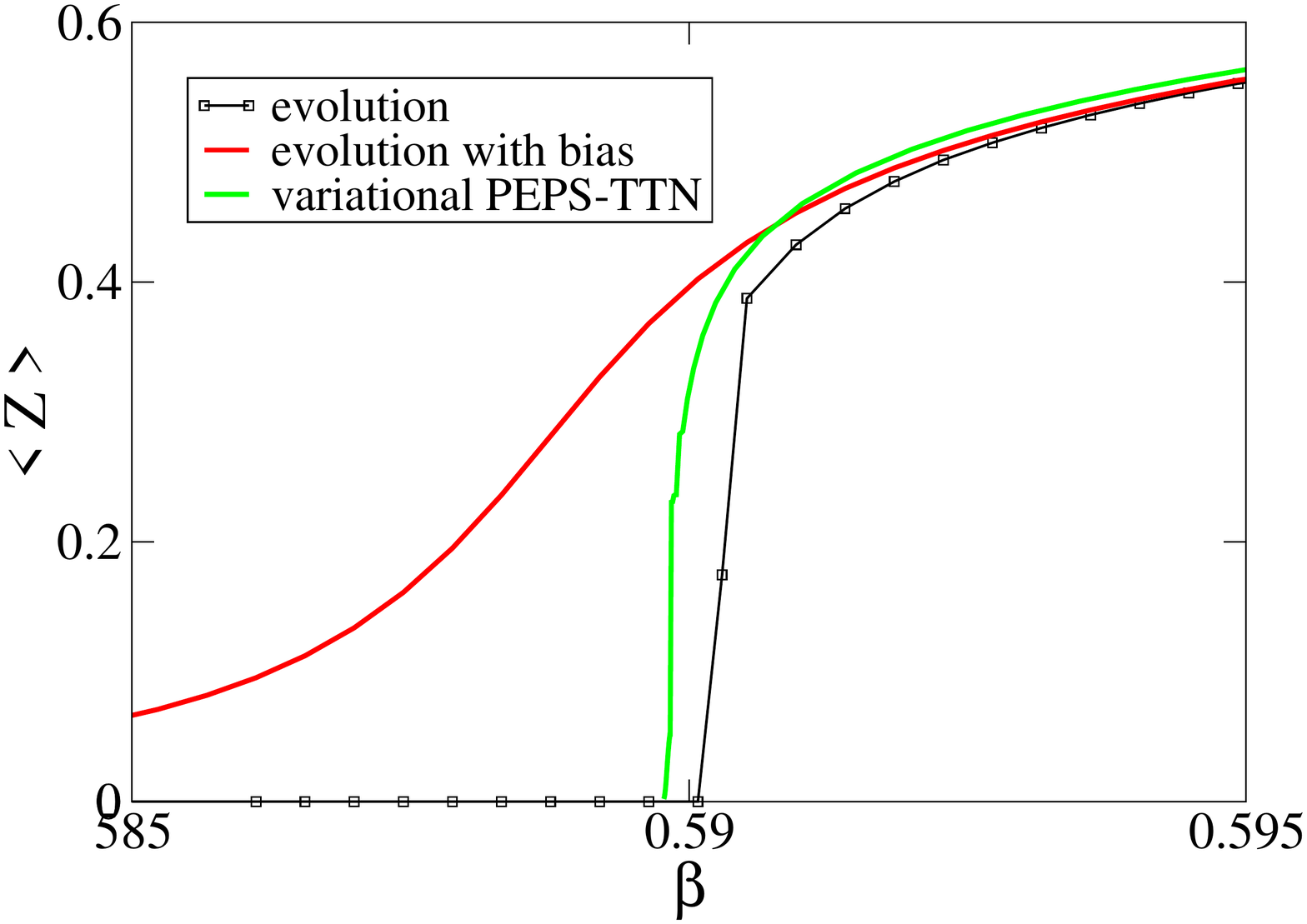}
\caption{ 
Spontaneous magnetization $\langle Z\rangle$ in function of $\beta$ obtained with different algorithms. 
The direct imaginary time evolution produces a plot with a discontinuity near the second order transition due to insufficient $M$. 
Adding a tiny symmetry breaking bias $\delta=10^{-6}$ smooths the transition making the necessary $M$ finite. 
Both evolution curves come from Ref. \cite{self}. 
They are compared with the PEPS-TTN curve from Fig. \ref{fig:U2dcrit}A. 
Here all the results for $D=6$.
}
\label{fig:compare}
\end{figure}

In our previous work \cite{Czarnik,self}, 
we made efforts to obtain thermal states by direct imaginary time evolution from $\beta=0$ to a finite one. 
In Fig. \ref{fig:ImTime} we compare our present method with the time evolution.
The most striking difference is the ansatz for the huge isometry $W$. 
In the time evolution
we apply one elementary Trotter tensor $T_0$ at every time step. Every application is followed
by a renormalization of the PEPS tensor with a new isometry $W_m$. This procedure implicitly assumes
the (left-canonical) matrix product state (MPS) ansatz in Fig. \ref{fig:ImTime}B. 
It has ${\cal O}\left(\beta/d\beta\right)$ different isometries, while in the corresponding TTN in 
Fig. \ref{fig:ImTime}A the same number is only ${\cal O}\left[\log(\beta/d\beta)\right]$. 
This is a dramatic difference for large $\beta$ or precise applications that require infinitesimal $d\beta$.

Another, possibly less apparent, difference is the optimization procedure. In the time evolution 
it is a one way process. After the $m$-th time step we choose the optimal isometry $W_m$ to renormalize 
the bond indices of the new PEPS tensor after this step. Once chosen, $W_m$ remains fixed and an error incurred 
with $W_m$ affects the whole following evolution. The errors accumulate with time. In the present variational 
approach we assume an entirely different strategy. Instead of scanning the whole range from $0$ to $\beta$ we 
target only the final $\beta$. The isometries are optimized by repeated up- and down-sweeps to provide the best 
final state. Each isometry alone and the whole set of isometries in tune are serving the single goal to make 
the targeted state as accurate as possible. Unlike in the time evolution, they are not compromised to provide 
accurate thermal states at intermediate imaginary times. Even if $T_n$ is the best PEPS tensor at the final 
$\beta$, $T_{n-s}$ does not need to be the best one at $\beta/s$.

The last property contributes to the main advantage of the variational PEPS-TTN over time evolution.
The variational algorithm can probe a low temperature phase without any need to evolve from infinite 
temperature across a critical point. Not only there is no direct evolution through intermediate temperatures, 
but also the intermediate tensors $T_{n-s}$ do not need to be thermal states at all.
The problems with evolution across the critical point are illustrated in Figure \ref{fig:compare}, 
where we compare three ferromagnetic magnetization curves obtained with three different algorithms.
The direct evolution runs into trouble at the critical point, where exact evolution would require
a divergent $M$, and the magnetization makes a discontinuous jump. This problem can be partly circumvented 
by adding a tiny symmetry breaking bias to the Hamiltonian,
\be 
\Delta H~=~-\delta\sum_m Z_m,
\ee 
that smooths the transition making the necessary $M$ finite, but significantly alters the physics at the critical
point. Nevertheless, the bias allows smooth evolution from infinite temperature deep into the low temperature 
phase, where the effect of the tiny bias becomes negligible. However, even the bias cannot prevent the
evolution from accumulating errors with time.

In retrospect,
it may be tempting to combine the new variational strategy with the MPS ansatz instead of TTN. 
After all, the powerful methods developed for MPS \cite{Schollwoeck} may be efficient enough to optimize
even the huge number ${\cal O}\left(\beta/d\beta\right)$ of isometries. Unfortunately, these methods
cannot be applied in 2D. Instead, the isometries in MPS have to be optimized by the procedures in 
Fig. \ref{fig:peps} combined with CMR. Since an isometry in MPS maps from $2D$ to $D$ dimensions, rather than from $D^2$ 
to $D$ in TTN, the procedures are more efficient for MPS than for TTN. However, the actual bottleneck that limits $D$ 
on an infinite lattice is the CMR whose cost depends only on the net $D$ and not on the underlying ansatz. Thus with 
MPS one can achieve the same $D$ as with TTN, but at the expense of an algorithm that is linear instead of logarithmic 
in $\beta/d\beta$. Apart from this, the linear algorithm has a lot more variational parameters, hence in principle it 
may be more liable to getting trapped in local maxima of the figure of merit. However, this discussion does not quite 
exclude MPS in some applications like, e.g., a finite lattice.   

\section{Conclusion}
\label{sec:conclusion} 
The proposed PEPS-TTN method is the first 2D finite-temperature tensor-network algorithm 
that is both variational and becomes numerically exact with increasing bond dimension. 
It avoids the demanding direct imaginary time evolution and employs full tensor 
environment in variational optimization. There is still a lot of room for improvement
and development. For instance, the simple TTN could be made more powerful with some unitary 
disentanglers \cite{MERA}. Internal symmetries, like $Z_2$ or $U(1)$, could help 
to use the bond dimension more efficiently. Finally, CMR may be upgraded to an algorithm 
making more efficient use of the environmental bond dimension. 

\acknowledgements
This work was supported by the Polish National Science Center (NCN) under project DEC-2013/09/B/ST3/01603.



\appendix
\section{Corner matrix renormalization (CMR)\label{CMR}}

\begin{figure}[t]
\includegraphics[width=0.99\columnwidth,clip=true]{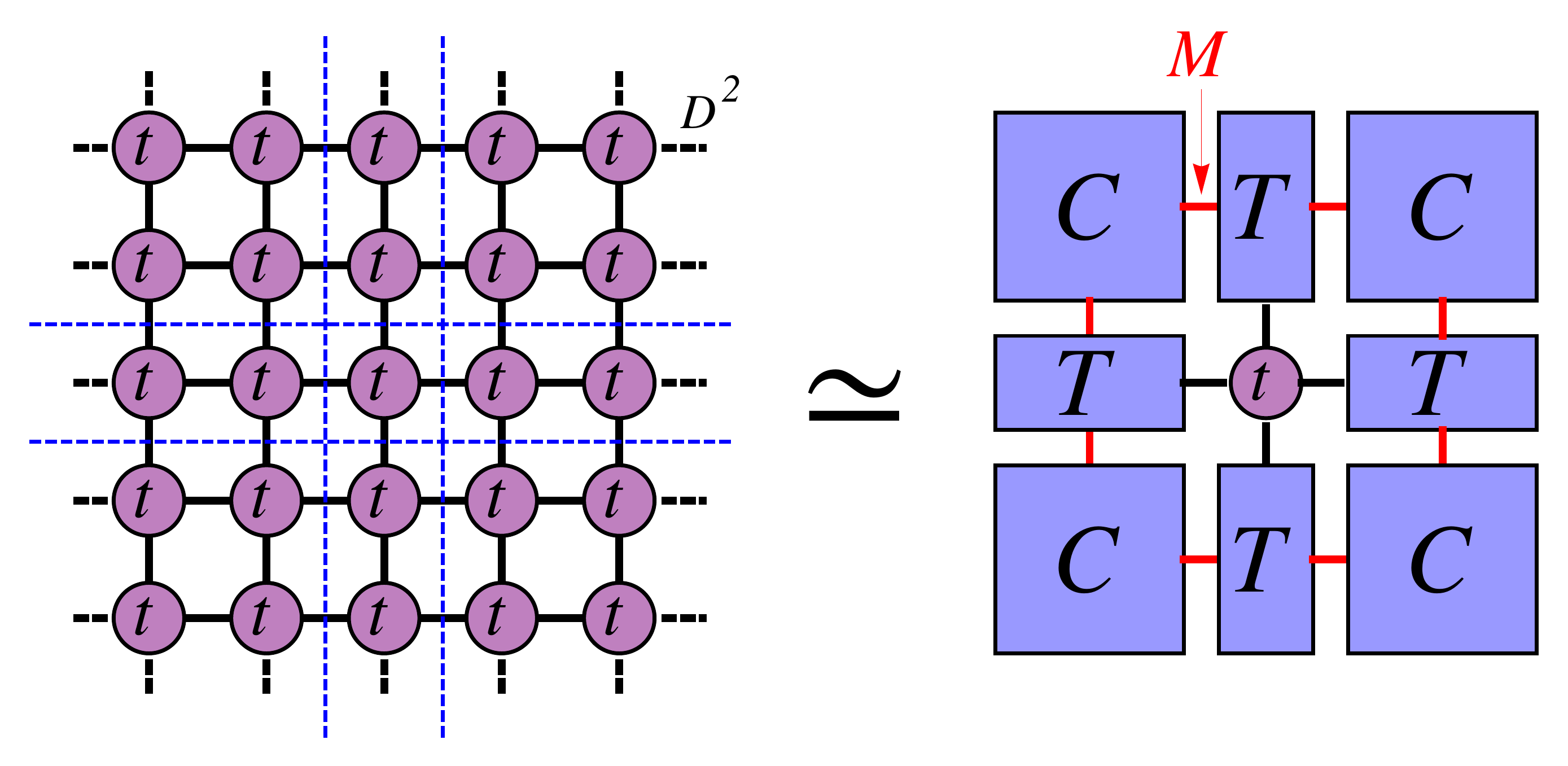}
\vspace{-0.2cm}
\caption{ 
On the left,
a planar version of the network in Fig. \ref{fig:peps}E representing the partition function.
Here each (black) bond represents two bond indices in Fig. \ref{fig:peps}E. 
Its dimension is $D^2$.
This infinite contraction cannot be done exactly,
hence it is approximated by the finite network on the right.
The finite corner matrices $C$ and top tensors $T$ effectively represent corresponding infinite sectors of 
the network on the left separated by the dashed blue lines. 
Their (red) environmental bonds have dimension $M$. 
The environmental tensors $C$ and $T$ should be such that
to the transfer tensor $t$ in the center 
its effective environment on the right appears the same as its exact environment on the left as much as possible.
They are obtained by iterating until convergence the corner matrix renormalization in Fig. \ref{FigRenC}.
}
\label{FigCV}
\end{figure}

\begin{figure}[t]
\includegraphics[width=0.8\columnwidth,clip=true]{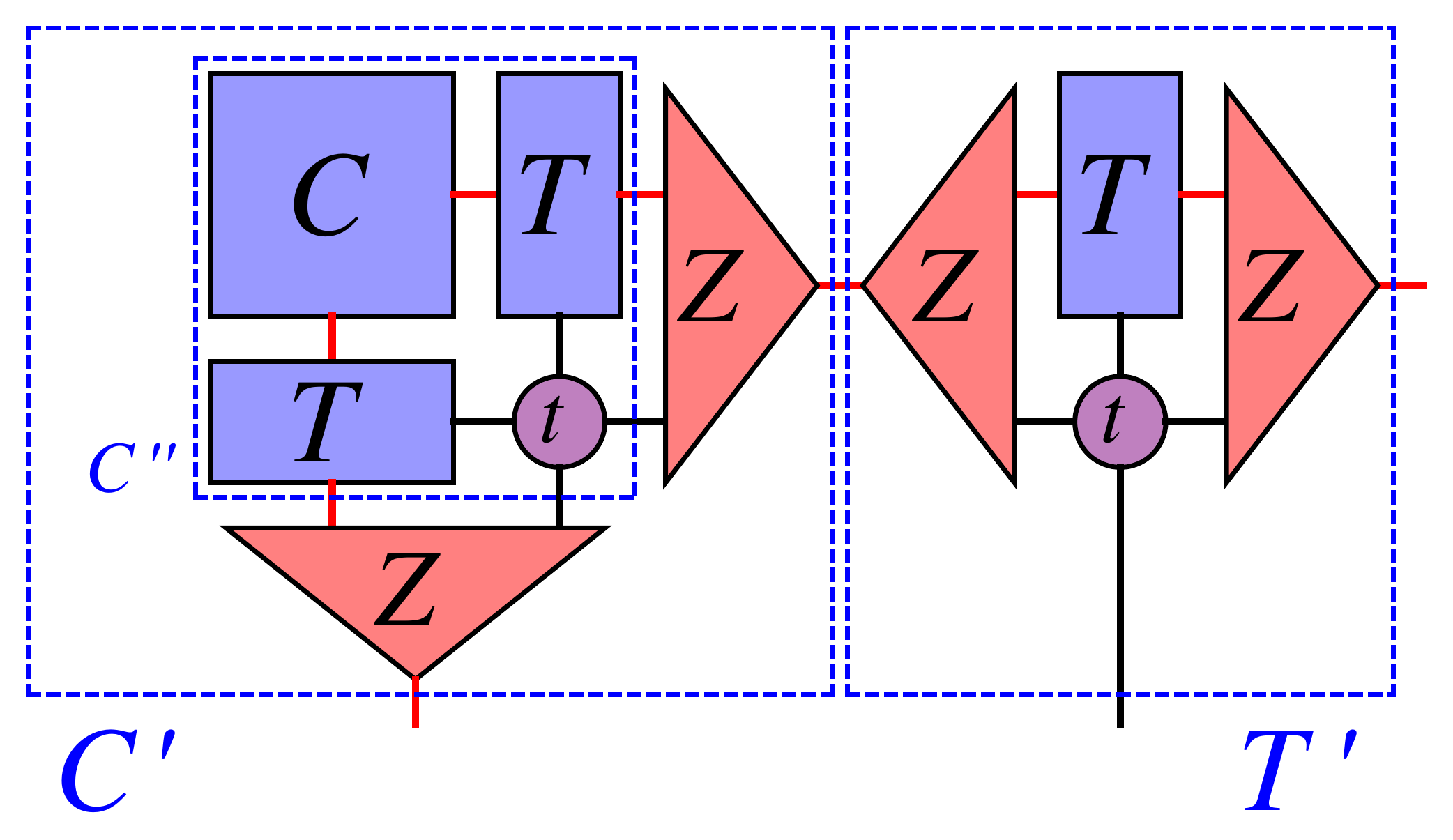}
\vspace{-0.2cm}
\caption{ 
The corner $C$ and top $T$ are obtained by repeating until convergence a renormalization procedure. 
The procedure has four steps. 
In the first step,
the tensors $C,T,t$ are contracted to an enlarged corner $C''$. 
In the second step,
the symmetric $MD^2\times MD^2$ matrix $C''$ is diagonalized and
its $M$ eigenvectors with the largest eigenvalues define an isometry $Z$. 
The diagonalization that scales like $M^3D^6$ is the leading cost
of this variant of corner matrix renormalization. 
In the third step,
$Z$ is used to renormalize/truncate the indices of $C''$ back to the original dimension $M$ 
giving a new (diagonal) corner $C'$.
In the fourth step, 
the same $Z$ renormalizes the contraction of $T$ with $t$ to a new $T'$. 
The four-step procedure is repeated until convergence of the $M$ leading eigenvalues of $C''$.
}
\label{FigRenC}
\end{figure}

\begin{figure}[t]
\includegraphics[width=0.8\columnwidth,clip=true]{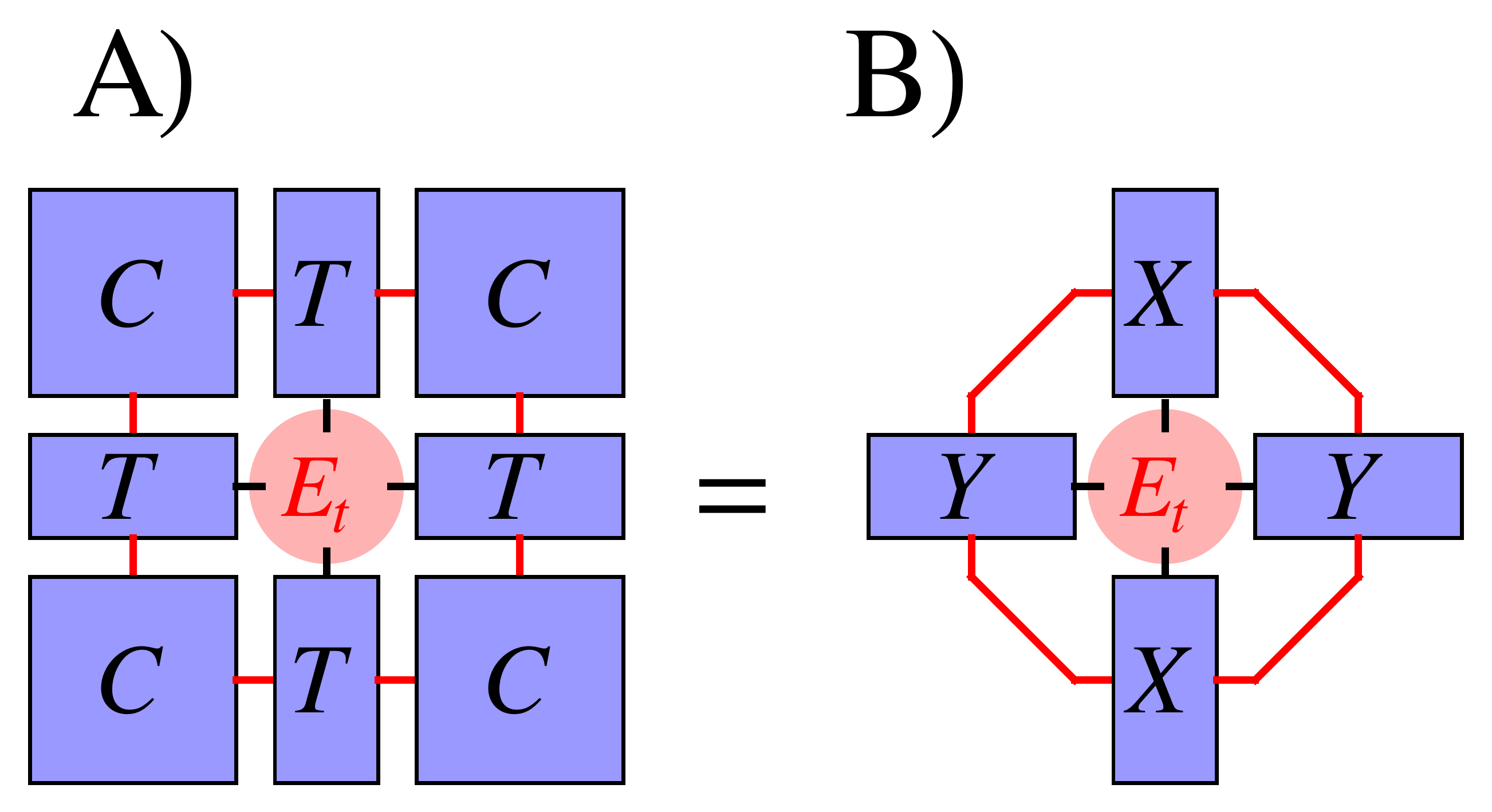}
\vspace{-0.2cm}
\caption{ 
In A,
a finite tensor network approximating the infinite environment $E_t$ in Fig. \ref{fig:peps}F.
Notice that the corner matrix $C$ can be singular-value-decomposed, $C=U\lambda V^\dag$,
and then absorbed into the neighboring top tensors $T$. With $X=\sqrt{\lambda}U^\dag TU\sqrt{\lambda}$ and $Y=\sqrt{\lambda}V^\dag T V\sqrt{\lambda}$ 
(here the matrix products are understood in the environmental bond indices) we obtain an equivalent network in panel B. 
This network is a matrix product state (MPS) of four tensors $X,Y,X,Y$ with a bond dimension $M$,
hence in principle a finite $M$ could suffice to represent the exact $E_t$ even at the critical point. 
}
\label{FigCTE}
\end{figure}

An infinite tensor network, 
like the one in Fig. \ref{fig:peps}E, 
cannot be contracted exactly.
Fortunately,
what we need in general is not this number,
but an environment for a few tensors of interest.
For instance, 
in Fig. \ref{fig:peps}F we want an environment $E_t$ for the transfer tensor $t$ in the center.
The environment is a tensor that remains after removing the central tensor $t$ from the infinite network.
From the point of view of the central tensor,
its infinite environment can be substituted with a finite effective environment, 
made of finite corner matrices $C$ and top tensors $T$,
that appears to the central tensor the same as the exact environment as much as possible.
The environmental tensors are contracted with each other by indices of dimension $M$.   
Increasing $M$ makes the effective environment more accurate and, 
for a finite correlation length, 
the effective environment is expected to converge to the exact one at a finite $M$. 
In the Ising model, 
tensor $t$ is symmetric under permutation of its indices and, 
consequently, 
$C$ is a symmetric matrix and $T$ is symmetric in its environmental indices.

Finite tensors $C$ and $T$ represent infinite sectors of the network on the left of Fig. \ref{FigCV}. The tensors are converged by iterating the corner matrix renormalization in Fig. \ref{FigRenC}. In every renormalization step, the corner matrix is enlarged with one $t$ and two $T$'s. This operation represents the 
top-left corner sector in Fig. \ref{FigCV} absorbing one more layer of tensors $t$. Once the environment is converged, it can be used to calculate either an observable or the environment $E_t$, see Fig. \ref{FigCTE}.

The above CMR procedure requires $M$ that diverges at a critical point. This is clearly demonstrated in the appendix of Ref. \cite{self} at the Onsager transition in the 2D classical Ising model, where a finite $M$ results in a finite correlation length $\propto M^{1.93}$. The tail of the correlation function cannot become strictly algebraic for any finite $M$ but, as illustrated in section \ref{sec:PEPS} and Ref. \cite{self}, with increasing $M$ not only local observables but also correlations at increasingly long distance become converged in $M$. At a critical point the effective environment is not exact, but provides an approximation whose quality improves with $M$ in a systematic way.   

In an attempt to go beyond this bottom line, one could argue that the divergent $M$ required at criticality is not an inherent property of the finite effective environment, but an artifact of the specific CMR method used to obtain this environment. Indeed, the exact tensor $E_t$ in Fig. \ref{fig:peps}F can be interpreted as a quantum state on the four free bonds, each free bond with $D^2$ auxiliary states. Such a finite state can be represented by a four-site MPS with a finite bond dimension $M$. In Fig. \ref{FigCTE}B we show that this MPS is equivalent to a finite effective environment with the same $M$. This completes the argument that for a finite $D$ the exact $E_t$ can be represented with a finite $M$.   

An exact local $E_t$ is all we need to optimize the isometries $W_m$ of the thermal PEPS, even though the finite $M$ of its exact environmental tensors may prohibit accurate calculation of the power-law tails of critical correlations. The present CMR method attempts to be universal -- accurate for both local and non-local observables -- while in order to optimize the PEPS tensor all we need is a method targeting the local $E_t$ only. 
  
\end{document}